\documentclass[journal]{IEEEtran}

\usepackage[cmex10]{amsmath}
\usepackage{amsmath,amssymb,amsfonts}
\usepackage{algorithm}
\usepackage{algorithmic}
\usepackage{graphicx}
\usepackage{textcomp}
\usepackage{bm}
\usepackage{setspace}
\usepackage{caption2}

\begin{document}

\title{$L_2$-Box Optimization for Green Cloud-RAN via Network Adaptation\\
	\thanks{Fan Zhang, Qiong Wu, Hao Wang, and Yuanming Shi are with School of Information Science and Technology, ShanghaiTech University, Shanghai, P.R. China (email: \{wuqiong, zhangfan4, wanghao1, shiym\}@shanghaiTech.edu.cn).}}%
\author{Fan Zhang,
	Qiong Wu,
	Hao Wang,
	and Yuanming Shi%
%The authors are with the Electronic and Computer Engineering Department, Technical University of Crete, 
%73132 Chania, Greece (e-mail: alex@telecom.tuc.gr, liavas@telecom.tuc.gr).}
}
%\author{Author 1 and Author 2}

\maketitle

\begin{abstract}
In this paper, we propose a reformulation for the Mixed Integer Programming (MIP) problem into an exact and continuous model through using the $\ell_2$-box technique to recast the binary constraints into a box with an $\ell_2$ sphere constraint.
The reformulated problem can be tackled by a dual ascent algorithm combined with a Majorization-Minimization (MM) method for the subproblems to solve the network power consumption problem of the Cloud Radio Access Network (Cloud-RAN),
and which leads to solving a sequence of Difference of Convex (DC) subproblems handled by an inexact MM algorithm. After obtaining the final solution, we use it as the initial result of the bi-section Group Sparse Beamforming (GSBF) algorithm to promote the group-sparsity of beamformers, rather than using the weighted $\ell_1 / \ell_2$-norm. Simulation results indicate that the new method outperforms the bi-section GSBF algorithm by achieving smaller network power consumption, especially in sparser cases, i.e., Cloud-RANs with a lot of Remote Radio Heads (RRHs) but fewer users.
\end{abstract}
\begin{IEEEkeywords}
Cloud-RAN, $\ell_2$-box, DC problems, MM algorithms, dual ascent methods, inexact algorithms,  group-sparsity.
\end{IEEEkeywords}

\section{Introduction}

\IEEEPARstart{I}{t} Cloud Radio Access Network (Cloud-RAN) is a network architecture proposed to meet the explosive growth of mobile data traffic. An important problem of Cloud-RAN is the energy efficiency consideration, due to the increasing power consumption of a large number of Remote Radio Heads (RRHs) as well as the fronthaul links. We focus on the power consumption problem of green Cloud-RAN by jointly involving the power consumption of the transport network and RRHs. Several methods have been proposed to solve the Cloud-RAN power minimization problem. 

The network power minimization problem can be formulated into a Mixed Integer Programming (MIP) problem, which is mainly solved by  three strategies. First of all, a global optimal solution can achieve by the branch-and-bound method~\cite{cheng2013joint}, but it may suffer from an exponential worst-case complexity and work slowly in practice. In order to alleviate the computational burden, Yang et. al.~\cite{cheng2013joint} derives an approximation of the MIP problem by relaxing the binary constraint to a $[0, 1]$ box constraint. %However, this method sacrifices the accuracy of solution. 
The most related method is a three-stage approach named Group Sparse Beamforming (GSBF) algorithm~\cite{Hong2013Joint, Dai2017Sparse}, which balances between the computational complexity and the accuracy of solution. This algorithm exploits the group sparsity structure of beamformers with the priori knowledge. Specifically, in the first stage, it solves a convex weighted $\ell_1/\ell_2$ norm relaxation of the MIP problem to induce the sparsity of the beamformers. The second stage generates an ordering rule to decide which RRH has a higher priority to be switched off. In the third stage, a selection procedure is performed to determine the best combination of the active and the sleep set of RRHs. However, the GSBF algorithm generally can not guarantee to provide a high accuracy solution.

In this paper, we propose a new formulation of the Cloud-RAN power consumption problem along with a dual ascent method combined with an inexact Majorization-Minimization (MM) algorithm. The major idea of this recast is the $\ell_2$-box technique, introduced recently in~\cite{wu2016ell_p} as a continuous equivalent formulation of the binary constraints. By using this technique to replace the binary constraint with the intersection of a box and $\ell_2$ sphere, we obtain a new formulation of the Cloud-RAN power consumption problem.
As a result, a local optimal solution can be quickly found by continuous algorithms, while for the original mixed-binary problem,  excessive computational effort may be needed to find a comparable solution. Therefore, the solution of the proposed reformulated problem can be employed to initialize the bi-section GSBF algorithm, which can be a more powerful sparsity-promoting tool than the weighted $\ell_1/\ell_2$ norm relaxation.
It should be emphasized that our exact and continuous formulation of the network power consumption problem, in contrast to a relaxation model, can enable a better solution be reached. 

The reformulated problem can be addressed by our proposed MM dual ascent algorithm and test by the numerical experiments. The numerical results manifest that our proposed framework obviously improves the network energy efficiency, especially in the case of more RRHs but fewer users.

\section{PROBLEM DESCRIPTION}
\subsection{System model}
We consider a Cloud-RAN with $L$ RRHs and $K$ single-antenna Mobile Users (MUs), where the $l$-th RRH is equipped with $N_l$ antennas. In this architecture, all the Baseband Units (BBU) are moved in to a single BBU pool, creating a set of shared processing resources, and enabling efficient interference management and mobility management. All the RRHs are connected to the BBU pool through fronthaul links. In a beamforming framework, let $\bm{v}_{lm} \in \mathbb{C}^{N_l}$ be the transmit beamforming vector from the $l$-th RRH to the $k$-th user, and $s_k$ be the data symbol for user $k$ with $\emph{E}[\left |  s_k\right |^2 =1]$. The transmit signal at RRH $l$ is given by
\begin{equation}
\bm{x}_{l}=\sum_{k=1}^{K}\bm{v}_{lm}s_k,\ \forall l \in \mathcal{L}
\label{eq1}.
\end{equation}
The channel propagation between user $k$ and RRH $l$ is denoted as $\bm{h}_{lm}\in \mathbb{C}^{N_l}$, and $n_k \in \mathcal{CN}(0, \sigma_k^2)$ is the additive Gaussian noise at user $k$. Therefore, the received signal at user $k$ is then
\begin{equation}
y_k = \sum _{l \in \mathcal{L}}\bm{h}_{kl}^{\sf{H}}\bm{v}_{lk}s_k+\sum_{i\neq k}\sum_{l \in \mathcal{L}}\bm{h}_{kl}^{\sf{H}}\bm{v}_{li}s_{i}+n_k
\label{eq2}.
\end{equation}

We assume that all the users treat the interference as noise~\cite{cadambe2008interference}. The corresponding signal-to-interference-plus-noise ratio (SINR) for user $k$ is 
\begin{equation}
\Gamma_k = \frac{\left | \sum _{l \in \mathcal{L}}\bm{h}_{kl}^{\sf{H}}\bm{v}_{lk} \right |^2}{\sum_{i\neq k}\left | \sum_{l \in \mathcal{L}}\bm{h}_{kl}^{\sf{H}}\bm{v}_{li} \right |^2+\sigma _k^2}.
\label{eq3}
\end{equation}
Each RRH has its own transmit power constraint
\begin{equation}
\sum_{k=1}^{K} \| \bm{v}_{lk}  \|_{2}^2 \leq \sqrt{P_l},\ \forall l \in \mathcal{L},
\label{eq4}
\end{equation}
where $P_l$ is the maximum transmit power of the $l$-th RRH.

\subsection{Network power consumption minimization}
Due to the high density of RRHs and their joint transmission, the energy used for signal transmission can be reduced significantly. However, the power consumption of the transport network becomes numerous and cannot be ignored. In order to reduce the network power consumption, it is essential to put some RRHs into sleep whenever possible. We introduce a binary vector $\bm{z} = (z_1,..., z_L)^T$ to represent the active RRH, i.e., $z_l = 1$ denotes the $l$-th RRH is active, and $z_l = 0$ means the $l$-th RRH is sleeping.
Denote the relative fronthaul link power consumption by $P_l^c$, and the inefficient of drain efficiency of the radio frequency power amplifier by $\eta_l$. Then the network power consumption $p (\bm{z}, \bm{v})$ is the sum of total transmit power consumption and the total relative fronthaul links power consumption:
\begin{equation}
p (\bm{z}, \bm{v}) = \sum_{l \in \mathcal{L}}P_l^c z_l + \sum_{l \in \mathcal{L}}\frac{1}{\eta_l} \|  \bm{\tilde{v}}_{l}  \|_2^2.
\label{eq5}
\end{equation}
where, for convenience, let $\bm{\tilde{v}}_{l}=[\bm{v}_{l1}^T,..., \bm{v}_{lK}^T]^T\in \mathbb{C}^{KN_l\times 1}$. With target SINRs $\bm{\gamma} = (\gamma_1,..., \gamma_K)^T$, the SINR constraint for user $k$ as a second-order cone (SOC) constraint~\cite{shi2014group} must be satisfied. Therefore, the power minimization problem can be formulated as a MIP problem~\cite{lee2011mixed}
\begin{equation}
\begin{aligned}
& \underset{(\bm{z},\bm{v})}{\operatorname{min}}
& & p (\bm{z}, \bm{v}) \\
& \text{s.t.}
& & \sqrt{\sum_{i\neq k} \| \bm{h}_k^{\sf{H}} \bm{v}_i \|_2^2 + \sigma _k^2}\leq \frac{1}{\gamma _k}\Re (\bm{h}_k^{\sf{H}} \bm{v}_k), k \in \mathcal{S},\\
&&&  \| \bm{\tilde{v}}_{l}  \| \leq z_l \sqrt{P_l}, z_l = \{0, 1\}  , l \in \mathcal{L} ,
%&&& l \in \mathcal{L} , k \in \mathcal{S},
\end{aligned}
\label{eq6}
\end{equation}
where $\Re(\cdot)$ denotes the real part. 

\section{$\ell_2$-box optimization reformulation}

In this section, we propose a new formulation of the Cloud-RAN power consumption problem. An $\ell_2$-box technique is proposed to replace the binary by the intersection between a box and an $\ell_2$ sphere as described in~\eqref{eq7}. 
A geometric illustration of the $\ell_2$-box technique is depicted in Fig. 1.
\begin{equation}
\bm{x} \in \{0, 1\}^n \Leftrightarrow \bm{x} \in [0, 1]^n \cap \left\{\bm{x}: \| \bm{x}- \frac{1}{2} \bm{\emph{1}}_n \|_2^2 = \frac{n}{4} \right \},
\label{eq7}
\end{equation}
\begin{figure}[t!]
	\centering
	\centerline{\includegraphics[width=0.4\linewidth]{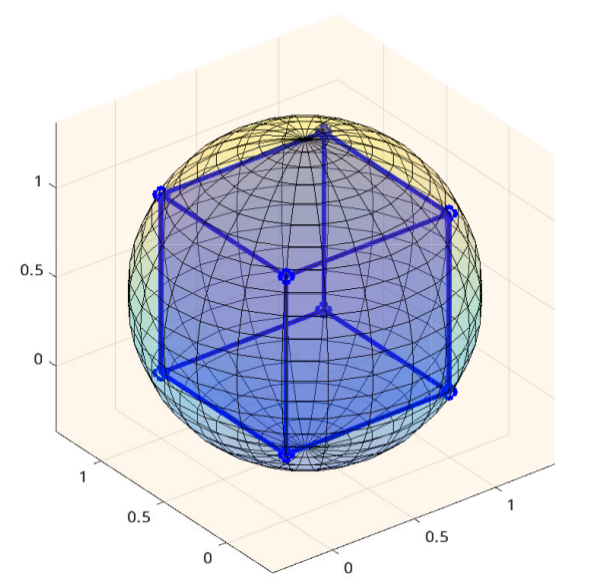}}
	%  \vspace{2.0cm}
	%\centerline{(a) Result 1}\medskip
	\caption{Geometric illustration of the $\ell_2$-box technique. The hollow circle is the intersection of the box and $\ell_2$-norm.}
	\label{fig:lpbox}
\end{figure}
where $\bm{\emph{1}}_n$ is an $n$-dimension all-one vector. Therefore, the MIP~\eqref{eq6} can be recast into
\begin{equation}
\begin{aligned}
& \underset{(\bm{z},\bm{v})}{\operatorname{min}}
& & \sum_{l \in \mathcal{L}}P_l^c z_l + \sum_{l \in \mathcal{L}}\frac{1}{\eta_l} \|  \bm{\tilde{v}}_{l}  \|_2^2  \\
& \text{s.t.}
& & \sqrt{\sum_{i\neq k} \| \bm{h}_k^{\sf{H}} \bm{v}_i \|_2^2 + \sigma _k^2}\leq \frac{1}{\gamma _k}\Re (\bm{h}_k^{\sf{H}} \bm{v}_k),  k \in \mathcal{S} ,\\
&&&  \| \bm{\tilde{v}}_{l}  \| \leq z_l \sqrt{P_l}, 0 \leq z_l \leq 1  , l \in \mathcal{L},\\
&&&  \| \bm{z} - \frac{1}{2}\bm{\emph{1}}_L  \|_2^2 = \frac{L}{4}.
%&&& l \in \mathcal{L} , k \in \mathcal{S}.
\end{aligned}
\label{eq8}
\end{equation}
%By applying the $\ell_2$-box technique, we avoid solving the intractable binary programming problem, and address a continuous problem instead. 
The main difficulty of this problem comes from the nonconvexity of the sphere constraint, and we are well aware that it may be effort-consuming to  find the global optimal solution. However, rather than solve~\eqref{eq8} directly for the global optimal solution, rather than use the global nonlinear method to solve~\eqref{eq8} until global optimal, we only use local nonlinear algorithm to address~\eqref{eq8} and use the (local) solution as the initial point in the first stage of the bi-section GSBF algorithm, which can further induce the group sparsity of the beamformers.  
%we replace the weighted mixed $\ell_1/\ell_2$-norm with adopting the new formulation to induce the group-sparsity of the beamformers as the first stage of the bi-section GSBF algorithm. We emphasize that an equivalent formulation of the network power consumption problem rather than the convex relaxation is solved, which can enable us find a more accuracy solution.

\section{MM Dual ascent Algorithm}
\label{section4}
In this section, we design a dual ascent algorithm~\cite{boyd2011distributed} incorporated with an inexact MM algorithm to solve our proposed $\ell_2$-box Cloud-RAN power minimization problem.
%algorithm framework of tackling the $\ell_2$-box problem is described, and it is denominated as the MM dual ascent algorithm. 
Notice that~\eqref{eq8} is a convex problem except for the nonconvex $\ell_2$ sphere constraint. Therefore, we focus on dealing with the sphere constraint to construct our algorithm. For simplicity, let
\begin{equation}
\phi(\bm{v},\bm{z}) = \sum_{l \in \mathcal{L}}P_l^c z_l + \sum_{l \in \mathcal{L}}\frac{1}{\eta_l} \|  \bm{\tilde{v}}_{l}  \|_2^2 ,
\label{eq9}
\end{equation}
and $\Omega = \{(\bm{z},\bm{v})|\sqrt{\sum_{i\neq k} \| h_k^{\sf{H}} \bm{v}_i \|_2^2 + \sigma _k^2}\leq \frac{1}{\gamma _k}\Re (h_k^{\sf{H}} \bm{v}_k), k \in \mathcal{K} ;\\
\| \bm{\tilde{v}}_{l}  \| _2\leq z_l \sqrt{P_l},0 \leq z_l \leq 1,l \in \mathcal{L} \}.$ Notice that $\Omega$ is a convex set. Now~\eqref{eq8} can be stated as
\begin{equation}
\begin{aligned}
& \underset{{(\bm{z},\bm{v})\in \Omega}}{\operatorname{min}}
& & \phi(\bm{z},\bm{v}) \\
& \text{s.t.}
&&  \| \bm{z} - \frac{1}{2}\bm{\emph{1}}_L  \|_2^2 = \frac{L}{4}. \\
\end{aligned}
\label{eq10}
\end{equation}
A natural way to solve such a problem is to dualize the sphere constraint. Letting $\lambda$ be the multiplier associated with the sphere constraint, the Lagrangian of~\eqref{eq10} is defined as
\begin{equation}
L(\bm{z},\bm{v},\lambda) = \phi(\bm{z},\bm{v})+\lambda(\frac{L}{4}- \| \bm{z} - \frac{1}{2}\bm{\emph{1}}_L  \|_2^2),\\
\label{eq11}
\end{equation}
for $(\bm{z},\bm{v})\in \Omega$.
An alternative option is to use the augmented Lagrangian, but we do not suggest such approach since it will severely increase the nonlinearity of the resulted subproblems by introducing a fourth-order polynomial in the objective.
The dual of objective is then given by~\eqref{eq12} 
\begin{equation}
g(\lambda) = \inf_{(\bm{z},\bm{v})\in \Omega} L(\bm{z},\bm{v},\lambda),
\label{eq12}
\end{equation}
and we have the dual problem~\eqref{eq13}
\begin{equation}
\max_{\lambda} g(\lambda) = \max_{\lambda}\inf_{(\bm{z},\bm{v})\in \Omega}L(\bm{z},\bm{v},\lambda).
\label{eq13}
\end{equation}
Now we are ready to provide our dual ascent framework. The dual ascent method consists of two stages: the first stage is to update the primal variables by minimizing the Lagrangian for a fixed dual variable $\lambda$,
\begin{equation}
(\bm{z}^{t+1},\bm{v}^{t+1}) = \text{arg}\min_{(\bm{z},\bm{v})\in \Omega} L(\bm{z},\bm{v},\lambda^t),
\label{eq14}
\end{equation}
and then update the dual variable based on the constraint residual
\begin{equation}
\lambda^{t+1} = \lambda^t + \alpha^t(\frac{L}{4}- \| \bm{z}^{t+1} - \frac{1}{2}\bm{\emph{1}}_L  \|_2^2),
\label{eq15}
\end{equation}
where $\alpha^t>0$ is the step size of the dual update.

Since $\bm{z}$ is restricted in $\Omega$, it holds true $\bm{z}^{t+1}\in [0, 1]^L$. It follows that $(\frac{L}{4}- \| \bm{z}^{t+1} - \frac{1}{2}\bm{\emph{1}}_L  \|_2^2) \geq 0$, where the equality holds true if and only if $\bm{z} \in \{0, 1\}^L$. As a result, the dual variable $\lambda^0 $ should be initialized to be positive to penalize the violation of the sphere constraint. Since the step size $\alpha^t > 0$, $\lambda$ is maintaining positive and increasing incrementally during the solution. 
Consequently, $\lambda^t \| \bm{z} - \frac{1}{2}\bm{\emph{1}}_L  \|_2^2$ is kept convex with respect to $\bm{z}$. In other words, the subproblem~\eqref{eq14} is a DC problem~\cite{hartman1959functions}. At the $s$-th iteration of MM algorithm~\cite{sun2017majorization}, a convex surrogate objective $\hat{L}(\bm{z},\bm{v},\lambda^t,\bm{z}^{(s)})$ is generated by linearizing the second convex function while keeping the first function unchanged
\begin{equation}
\begin{aligned}
\hat{L}(\bm{z},\bm{v},\lambda^t,\bm{z}^{(s)}) = \phi(\bm{z},\bm{v})+  
\lambda^t(\frac{L}{4}- \| \bm{z}^{(s) }- \frac{1}{2}\bm{\emph{1}}_L  \|_2^2\\
-2(\bm{z}-\frac{1}{2}\bm{\emph{1}}_L)^T(\bm{z}-\bm{z}^{(s)})),
\label{eq16}
\end{aligned}
\end{equation}
where $(\bm{z},\bm{v})\in \Omega$ and $\{( \bm {z^{(s)}}, \bm{v^{(s)}})\}$ represents a sequence of primal iterates for the subproblems. To obtain $( \bm {z^{(s+1)}}, \bm{v^{(s+1)}})$ in each iteration of MM algorithm, we need to solve
\begin{equation}
\min_{(\bm{z},\bm{v}) \in \Omega}\hat{L}(\bm{z},\bm{v},\lambda^t,\bm{z}^{(s)}),
\label{eq17}
\end{equation}
which can easily be solved by CVX solver~\cite{grant2008cvx}.
It should be noticed that generally the subproblem does not need to be solved exactly if sufficient improvement on the primal variable can be achieved. Therefore, we also solve~\eqref{eq14} inexactly, meaning we only solve a few subproblems~\eqref{eq17}. This inexact strategy has proven to be able to reduce the computational cost substantially
The description of the entire MM dual ascent algorithm is stated in Algorithm~\ref{alg.MMdual}.

The convergence analysis of dual ascent method is provided by~\cite{Bertsekas1999Nonlinear}.
Since MM algorithm is proposed~\cite{Kiers2016Discussion} as a generalization of the EM algorithm, MM algorithm inherits the convergence properties of the EM algorithm~\cite{vaida2005parameter}. The convergence results of the EM algorithm includes: the likelihood sequence of the EM algorithm is nondecreasing and convergent~\cite{Dempster1977Maximum}, and that the limit points of the EM algorithm are stationary points of the likelihood~\cite{Wu1983On}.

After obtaining the sparse beamformer $\bm{{v^*}}$, we use its group sparsity to generate the ordering criterion,
and then adopt the same binary search procedure as the bi-Section GSBF algorithm in~\cite{shi2014group} to obtain the final results.

\begin{algorithm}[htbp]
	\caption{MM dual ascent algorithm}
	\label{alg.MMdual}
	\begin{algorithmic}[1]
		\STATE Given the tolerances $\epsilon_1>0,\ \epsilon_2>0$ and $\epsilon_3>0$.
		\STATE Initialize $\bm{z}, \bm{v}$ and $\lambda>0$.
		%		\STATE Set $s$ = 0.  
		\WHILE{$| \lambda^{t+1}-\lambda^t  |\geq \epsilon_2$ or $\| \bm{z}^{t+1}-\bm{z}^t  \|_2 +
			\| \bm{v}^{t+1}-\bm{v}^t  \|_F \geq \epsilon_3$}
		\WHILE{$ \| \bm{z}^{(s+1)}-\bm{z}^{(s)}  \|_2 +
			\| \bm{v}^{(s+1)}-\bm{v}^{(s)}  \|_F\geq \epsilon_1$}
		\STATE \textbf{Update $(\bm{z},\bm{v})$}:\\
		$(\bm{z}^{(s+1)},\bm{v}^{(s+1)}) = \text{arg}\min_{\bm{z},\bm{v}} \hat{L}(\bm{z},\bm{v},\lambda^t,\bm{z}^{(s)})
		$
		\STATE $s = s+1$
		\ENDWHILE
		\STATE $\bm{z}^t = \bm{z}^{(s)}, \bm{v}^t = \bm{v}^{(s)}$
		\STATE \textbf{Update $\lambda$}:\ $\lambda^{t+1} = \lambda^t + \alpha^t(\frac{L}{4}- \| \bm{z}^t - \frac{1}{2}\bm{\emph{1}}_L  \|_2^2)$		
		\STATE  Set  $t = t+1$
		
		\ENDWHILE
		\STATE \textbf{return} $\bm{\tilde{v}}_{l}$, for $l = 1,...,L$.
	\end{algorithmic}
\end{algorithm}
\section{Simulation Results}

In this section, we describe the experimental setting including the initial point and the algorithm parameters. In our experiment we check the convergence of the proposed method, and exhibits the effectiveness of our proposed method compared with contemporary methods. 

The initial point plays a significantly important role while solving the nonconvex problems. Instead of randomly choosing initial point, we remove the sphere constraint and solve the approximation problem
\begin{equation}
(\bm{z}^0, \bm{v}^0) = \text{arg}\min_{(\bm{z},\bm{v})\in \Omega} \phi(\bm{z},\bm{v})
\label{eq18}
\end{equation}
to derive the initial point. This generally renders better estimate than random initial point of the solution. 
The DC subproblem is solved inexactly under the stopping criterion $\| \bm{z}^{t+1}-\bm{z}^t  \|_2 \leq \epsilon_1$ with $\epsilon_1 = 10^ {-5}$. The main algorithm is terminated whenever the primal iterates or the dual iterates converge, i.e., we use termination criterion $| \lambda^{t+1}-\lambda^t  |\leq \epsilon_2$ or $\| \bm{z}^{s+1}-\bm{z}^s  \|_2 + \| \bm{v}^{s+1}-\bm{v}^s  \|_F \leq \epsilon_3$ with $\epsilon_2 =10^{-2}, \epsilon_3 =10^{-3}$. 

In our experiment, we consider a network with $L = 10$, $K = 6$, $2$-antenna RRHs and single-antenna MUs uniformly and independently distributed in the square region $[-1000,\ 1000]\times[-1000,\ 1000]$ meters. We set all the relative transport link power consumption to be $P_l^c = 13W,\ l = 1, . . . , L$, and the inefficient of power amplifier~\cite{auer2011much} at each RRH is $\eta_l = \frac{1}{4}$.

\begin{figure}[hbt]
	\begin{minipage}[b]{0.495\linewidth}
		\centering
		\centerline{\includegraphics[width=4cm]{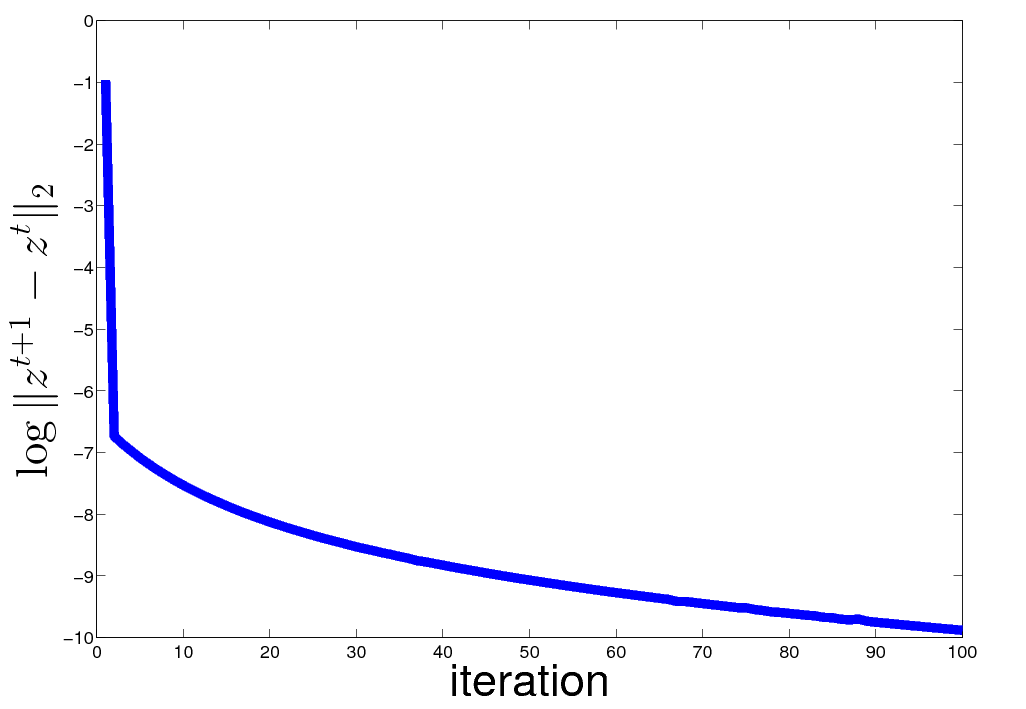}}
		\centerline{(a) Evolution of $ {tol}_1 $}\medskip
		\label{fig:side:a}
	\end{minipage}
	%	\begin{minipage}[b]{0.33\linewidth}
	%		\centering
	%		\centerline{\includegraphics[width=2.8cm]{z}}
	%		\centerline{(b) difference of $\bm{z}$}\medskip
	%		\label{fig:side:b}
	%	\end{minipage}%
	\begin{minipage}[b]{0.495\linewidth}
		\centering
		\centerline{\includegraphics[width=4cm]{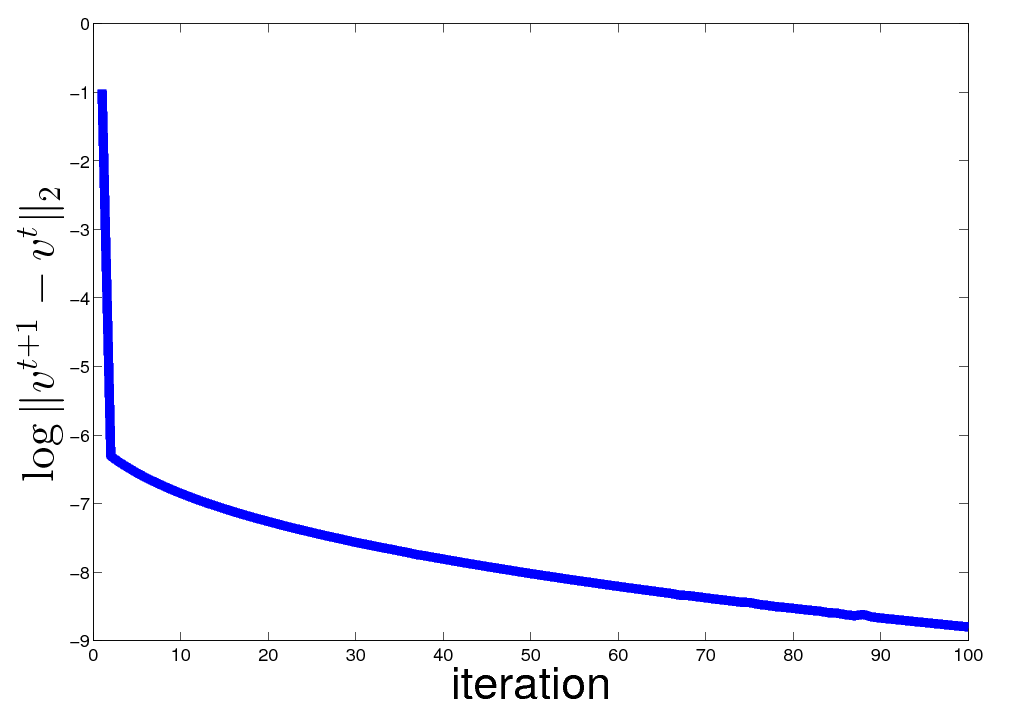}}
		\centerline{(b) Evolution of $ {tol}_2 $}\medskip
		\label{fig:side:b}
	\end{minipage}
	
	\caption{Convergence of the primal and dual variables.}
	\label{fig:convergence}
\end{figure}

In our first experiment, we show the efficiency of Algorithm~\ref{alg.MMdual}. The evolution of $ tol_1^t = \log{\| \bm{z}^{t+1}-\bm{z}^t  \|_2} $ and $ tol_2^t = \log{\| \bm{v}^{t+1}-\bm{v}^t  \|_F}$, where $\|\cdot\|_F$ is the Frobenius norm, is depicted in Fig.~\ref{fig:convergence}.(a) and (b) to represent the differences between the current and the previous iterates. 
%The change of $\lambda$ causes the ascent of primal variable $\bm{v}$ and $\bm{z}$. Therefore, we show the sequence of the change of primal variables just after the updating of $\lambda^t$ to test the convergence of primal variable. 
As shown in Fig.~\ref{fig:convergence}, both of the primal variables $\bm{z}$ and $\bm{v}$ are converging efficiency. In particular, $ {tol}_1 $ and $ {tol}_2$ decrease dramatically about $10^{-5}$ in the first five iterations.

\begin{figure}[t!]
	
	\begin{minipage}[b]{1.0\linewidth}
		\centering
		\centerline{\includegraphics[width=5.5cm]{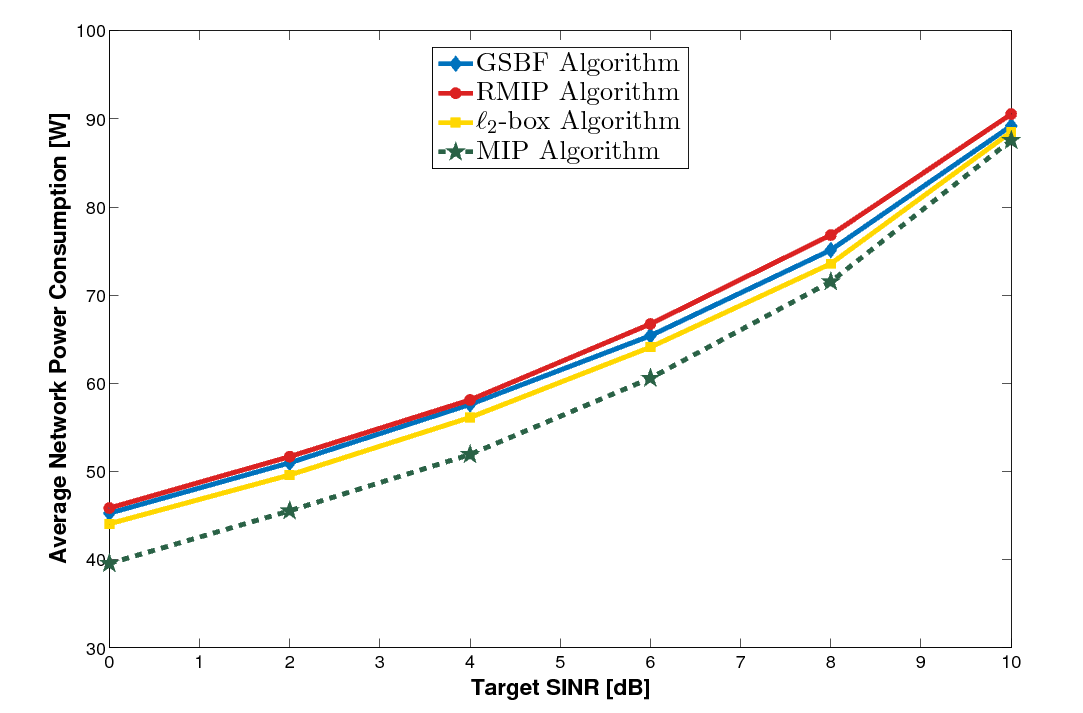}}
		%  \vspace{2.0cm}
		%\centerline{(a) Result 1}\medskip
	\end{minipage}
	
	\caption{Average network power consumption versus target SINR.}
	\label{fig:result}
\end{figure}

We also compare our proposed method with the existing methods including: MIP which is the branch-and-bound algorithm for solving the MIP problem~\eqref{eq6} for global optimal solution, RMIP which is the algorithm in~\cite{cheng2013joint} for solving the relaxed MIP problem, and GSBF which is the bi-section GSBF algorithm~\cite{Hong2013Joint, Dai2017Sparse}. 
The average network power consumption with different target SINR is shown in Fig.~\ref{fig:result}. The simulation results indicate that the $\ell_2$-box algorithm outperforms the GSBF and RMIP algorithm for different target SINR. This advantage becomes obvious in situations with smaller SINR.

\section{Conclusion and Future Work}
In this paper, we have proposed a new formulation of the Cloud-RAN power consumption problem by using the $\ell_2$-box technique, which replaces the binary constraint to two continuous constraints: a box constraint and a sphere constraint.
We design a dual ascent algorithm to solve the new $\ell_2$-box optimization problem leading a sequence of DC subproblems. We apply MM algorithm to inexactly solve the subproblem. The effectiveness of our proposed reformulation and algorithm is demonstrated in numerical experiment.
Our method exhibited lower network consumptions of different target SINR than the GSBF algorithm. 

Our investigation leads to a variety of open questions.
The final solution found by a nonlinear solver is often sensitive to the initial point. Therefore, it would be useful to explore better estimate of the global optimal solution to initialize our algorithm.
Furthermore, the binary constraint is also equivalent to the intersection of a box and an $\ell_p$ sphere with $p \in (0,+\infty)$. It would be interesting to investigate the performance of other $\ell_p$-box techniques, e.g., $\ell_1$-box or $\ell_{\frac{1}{2}}$-box. Moreover, besides dual ascent method, there are many other options for solving the proposed nonlinear problem, we leave it a subject for future work to investigate the performance other existing nonlinear solvers for our proposed problem.
	
\bibliographystyle{IEEEtran}
\bibliography{IEEEabrv,strings}

\begin{thebibliography}{10}
\providecommand{\url}[1]{#1}
\csname url@samestyle\endcsname
\providecommand{\newblock}{\relax}
\providecommand{\bibinfo}[2]{#2}
\providecommand{\BIBentrySTDinterwordspacing}{\spaceskip=0pt\relax}
\providecommand{\BIBentryALTinterwordstretchfactor}{4}
\providecommand{\BIBentryALTinterwordspacing}{\spaceskip=\fontdimen2\font plus
\BIBentryALTinterwordstretchfactor\fontdimen3\font minus
  \fontdimen4\font\relax}
\providecommand{\BIBforeignlanguage}[2]{{%
\expandafter\ifx\csname l@#1\endcsname\relax
\typeout{** WARNING: IEEEtran.bst: No hyphenation pattern has been}%
\typeout{** loaded for the language `#1'. Using the pattern for}%
\typeout{** the default language instead.}%
\else
\language=\csname l@#1\endcsname
\fi
#2}}
\providecommand{\BIBdecl}{\relax}
\BIBdecl

\bibitem{cheng2013joint}
Y.~Cheng, M.~Pesavento, and A.~Philipp, ``Joint network optimization and
  downlink beamforming for comp transmissions using mixed integer conic
  programming,'' \emph{IEEE Transactions on Signal Processing}, vol.~61,
  no.~16, pp. 3972--3987, 2013.

\bibitem{Hong2013Joint}
M.~Hong, R.~Sun, H.~Baligh, and Z.~Q. Luo, ``Joint base station clustering and
  beamformer design for partial coordinated transmission in heterogeneous
  networks,'' \emph{IEEE Journal on Selected Areas in Communications}, vol.~31,
  no.~2, pp. 226--240, 2013.

\bibitem{Dai2017Sparse}
B.~Dai and W.~Yu, ``Sparse beamforming and user-centric clustering for downlink
  cloud radio access network,'' \emph{IEEE Access}, vol.~2, pp. 1326--1339,
  2017.

\bibitem{wu2016ell_p}
B.~Wu and B.~Ghanem, ``$\ell_p $-box admm: A versatile framework for integer
  programming,'' \emph{arXiv preprint arXiv:1604.07666}, 2016.

\bibitem{cadambe2008interference}
V.~R. Cadambe and S.~A. Jafar, ``Interference alignment and degrees of freedom
  of the $ k $-user interference channel,'' \emph{IEEE Transactions on
  Information Theory}, vol.~54, no.~8, pp. 3425--3441, 2008.

\bibitem{shi2014group}
Y.~Shi, J.~Zhang, and K.~B. Letaief, ``Group sparse beamforming for green
  cloud-ran,'' \emph{IEEE Transactions on Wireless Communications}, vol.~13,
  no.~5, pp. 2809--2823, 2014.

\bibitem{lee2011mixed}
J.~Lee and S.~Leyffer, \emph{Mixed integer nonlinear programming}.\hskip 1em
  plus 0.5em minus 0.4em\relax Springer Science \& Business Media, 2011, vol.
  154.

\bibitem{boyd2011distributed}
S.~Boyd, N.~Parikh, E.~Chu, B.~Peleato, and J.~Eckstein, ``Distributed
  optimization and statistical learning via the alternating direction method of
  multipliers,'' \emph{Foundations and Trends{\textregistered} in Machine
  Learning}, vol.~3, no.~1, pp. 1--122, 2011.

\bibitem{hartman1959functions}
P.~Hartman, ``On functions representable as a difference of convex functions,''
  \emph{Pacific Journal of Mathematics}, vol.~9, no.~3, pp. 707--713, 1959.

\bibitem{sun2017majorization}
Y.~Sun, P.~Babu, and D.~P. Palomar, ``Majorization-minimization algorithms in
  signal processing, communications, and machine learning,'' \emph{IEEE
  Transactions on Signal Processing}, vol.~65, no.~3, pp. 794--816, 2017.

\bibitem{grant2008cvx}
M.~Grant, S.~Boyd, and Y.~Ye, ``Cvx: Matlab software for disciplined convex
  programming,'' 2008.

\bibitem{Bertsekas1999Nonlinear}
D.~P. Bertsekas, ``Nonlinear programming: 2nd edition,'' 1999.

\bibitem{Kiers2016Discussion}
H.~A.~L. Kiers, ``Discussion of article "optimization transfer using surrogate
  objective functions by lange, k. hunter, d.r. \& yang, i.",'' \emph{Journal
  of Computational \& Graphical Statistics}, vol.~9, 2016.

\bibitem{vaida2005parameter}
F.~Vaida, ``Parameter convergence for em and mm algorithms,'' \emph{Statistica
  Sinica}, pp. 831--840, 2005.

\bibitem{Dempster1977Maximum}
A.~P. Dempster, N.~M. Laird, and D.~B. Rubin, ``Maximum likelihood from
  incomplete data via the em algorithm,'' \emph{Journal of the Royal
  Statistical Society}, vol.~39, no.~1, pp. 1--38, 1977.

\bibitem{Wu1983On}
C.~F.~J. Wu, ``On the convergence properties of the em algorithm,''
  \emph{Annals of Statistics}, vol.~11, no.~1, pp. 95--103, 1983.

\bibitem{auer2011much}
G.~Auer, V.~Giannini, C.~Desset, I.~Godor, P.~Skillermark, M.~Olsson, M.~A.
  Imran, D.~Sabella, M.~J. Gonzalez, O.~Blume \emph{et~al.}, ``How much energy
  is needed to run a wireless network?'' \emph{IEEE Wireless Communications},
  vol.~18, no.~5, 2011.

\end{thebibliography}

\end{document}